\documentclass[twocolumn,secnumarabic,amssymb, nobibnotes, aps, prd]{revtex4-1}
\usepackage{graphicx, color}

\usepackage{amsmath}
\usepackage{booktabs}

\begin{document}
\title{Kinetics of nano-size Ferroelectrics}
\author{S. R. Etesami, A. Sukhov, and J. Berakdar}
\address{Institut f\"ur Physik, Martin-Luther-Universit\"at Halle-Wittenberg, 06099 Halle, Germany}
\date{\today}
\begin{abstract}
We analyze theoretically the finite-temperature polarization dynamic in displacive-type ferroelectrics.
 In particular we consider the thermally-activated switching time of a single-domain ferroelectric polarization studied by means  of the Landau-Khalatnikov equation,
 extended as to capture thermal fluctuations. The results are  compared with the switching time formula that follows from
  the analytical solution of Pauli master equations.
   In a second step we focus on the phase diagram of a prototypical ferroelectric as described by the temperature-dependent Landau-Devonshire model including  thermal fluctuations. Our simulations show the emergence of  phase instability at reduced sizes which we  attribute to  thermal fluctuations in the order parameter in the respective phase. We conclude that, along with the temperature-dependent potential coefficients, thermal fluctuations should  be taken into account to achieve a comprehensive description of the thermal behavior of reduced-size ferroelectrics. To endorse our conclusions, we simulated the electric-field activated switching time for a multi-domain system and compared the results  to the predictions of  well-established models such as the Kolmogorov-Avrami-Ishibashi.\\
\end{abstract}
\maketitle
\section{Introduction}
Thermal effects in displace-type ferroelectric (FE) materials are diverse. On the one hand for a broad range of temperatures, one observes  a variation in the polarization \cite{StLe98,Merz49,vHip50,Scott4} resulting from structural changes at certain transition temperatures. It is  common  to capture this behavior by means of the Landau-Devonshire potential with temperature-dependent coefficients \cite{Li,Wang,Wang1}. On the other hand  within a certain structural phase, another temperature-dependent dynamics is also operational which stems  from  the coupling of the order parameter to an external thermal bath and acts as a stochastic electric field \cite{LiAl07,KiLu13}. As shown in the present study the latter is particularly crucial at reduced-size FEs, in which case the thermal fluctuations can not be simply averaged out in the potential coefficients.
%In fact,  finite size effects in  FEs have been under investigation for decades
Among the finite-size effects in FEs \cite{Sedykh,LiEastman1996,Hoshina,Ishikawa,Lichtensteiger,Junquera,Lee,Scott,Zhao}, we address the suppression of the polarization at reduced sizes. This effect has been mainly attributed to the depolarization which arises from the uncompensated charges at the interfaces \cite{Scott,LiEastman1996}. We will show that thermal fluctuations may also play a key role in the occurrence of this phenomenon, and of the phase instability in general. To provide further evidence for the importance of thermal fluctuations, we simulated also the electric-field activated switching time in multi-domain systems within our formalism and found that the temperature-dependent potential coefficients alone cannot assist the polarization switching properly.
To elucidate how  thermal fluctuations are introduced into the models quantitatively, we follow the conventional Fokker-Planck approach. In Sec.~\ref{1} we do this and  evaluate the thermally activated switching time \cite{HuCh98,RiHa98,AhCa03,NgAh09,CaTa99,SiWi04,Klot08}  comparing the results with an already theoretically derived expression for the switching times  assuring the consistency of the two approaches.

\section{Thermally activated switching times}
\label{1}
\subsection{Analytical framework}
\begin{center}
   \begin{figure*}[t!]
    \centering$
        \begin{array}{c}
    \includegraphics[width=17cm]{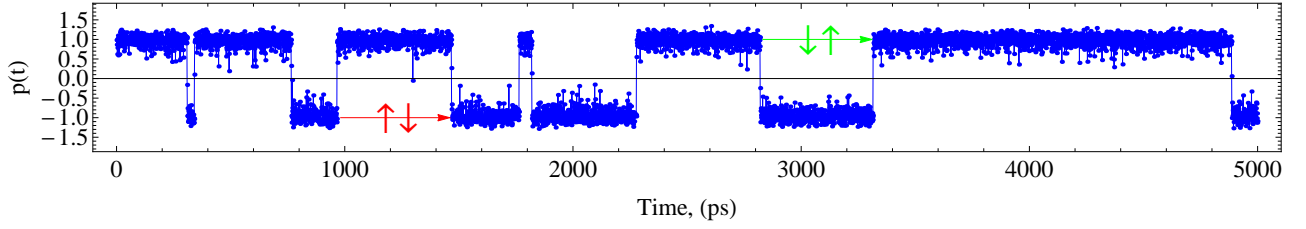}
        \end{array}$
        \caption{\label{p_time} \textbf{Thermally activated switching time :} Reduced polarization $p(t)=P(t)/P_S$ versus time, where $P_S=\sqrt{-\alpha/\beta}\approx0.74$ Cm$^{-2}$ is the spontaneous polarization. Via averaging over the red (green) time intervals $\tau_{\uparrow\downarrow}$ ($\tau_{\downarrow\uparrow}$) is evaluated. Since the polarization does not relax \emph{exactly} to the potential minima, even after switching (due to thermal fluctuations), the switching time is defined as the time in which the reduced polarization passes from $p>0.5$ to $p<-0.5$, and vise versa. For  more accurate results the simulation time interval $\Delta$ should be increased such that the averages are taken over a larger number of switchings. Further parameters are: $\Delta=5000$ ps, $a^3=10^{-27}$ m$^3$, $\gamma=25\times10^{-6}$ VmsC$^{-1}$, $T=200$ K and $E=1$ MVm$^{-1}$.}
    \end{figure*}
\end{center}
Let us start with a  simple model within the Landau-Devonshire framework for a single tetragonal FE crystal with the free energy density
\begin{equation}
\label{free_energy}
\begin{split}
F=\frac{\alpha}{2}P^2+\frac{\beta}{4}P^4-PE.
\end{split}
\end{equation}

The Landau coefficients are $\alpha=-11.57\times10^7$ VmC$^{-1}$, and $\beta=2.1\times10^8$ Vm$^5$C$^{-3}$. P and E are the polarization and the external electric field, respectively. In the absence of external electric field, the free energy has two symmetric minima which correspond to different stable orientations of the polarization, up($\uparrow$) and down($\downarrow$). At finite temperatures ($T$) there is a finite probability to find the polarization in either orientation. The averaged time in which the polarization stays in one orientation without switching to another one is called the switching time ($\tau$). Due to the kinetic energy attained by the system at finite temperatures, one expects  the switching time to decrease upon increasing $T$. In the presence of  an external electric field, the symmetry is broken and the probabilities for the polarization to be oriented up or down ($\Omega_\uparrow$ or $\Omega_\downarrow$) becomes different. Therefore, different switching times are expected ($\tau_{\uparrow\downarrow}$ and $\tau_{\downarrow\uparrow}$) (see Fig.~\ref{p_time}). Numerically one can compare the results with the formula from Ref. \cite{Vopsaroiu}
\begin{equation}
\label{Switching_time_theory}
 \tau^{-1}_{\downarrow\uparrow}=v_0\left (e^{W_\uparrow/k_BT}+e^{W_\downarrow/k_BT}\right ),
\end{equation}
which is the analog of the semi-empirical Kolmogorov-Avrami-Ishibashi (KAI) model \cite{Kolmogorov,Avrami,Ishibashi}. $W_{\uparrow,\downarrow}$ are the energies at the minima where for small electric fields ($E\ll\frac{4}{3}\sqrt{-\frac{\alpha^3}{\beta}}\approx114$ MVm$^{-1}$) are approximated as $W_{\uparrow,\downarrow}\approx a^3(-\frac{\alpha^2}{4\beta}\mp\sqrt{-\frac{\alpha}{\beta}}E)$. In Eq. (\ref{Switching_time_theory}) $v_0$ is the total number of trials per second to overcome the energy barrier between the minima.  Experimentally this corresponds to the frequency of the optical phonons. According to Ref. \cite{Cochran} $v_0$ is temperature-dependent with a  value being  determined by the distance to the Curie temperature. Since the Curie temperature is introduced via the Landau coefficients in the simulations, here we have chosen them temperature-independent, so that the temperature dependency of $v_0$ is dismissible. For a detailed  derivation of the switching time formula we refer to Ref.~\cite{Vopsaroiu}; here we briefly mention that it is derived from the Pauli master equations $\frac{d\Omega_\uparrow}{dt}=a_{\uparrow\downarrow}\Omega_\downarrow-a_{\downarrow\uparrow}\Omega_{\uparrow}$ and $\frac{d\Omega_\downarrow}{dt}=a_{\downarrow\uparrow}\Omega_\uparrow-a_{\uparrow\downarrow}\Omega_{\downarrow}$, where $a_{\uparrow\downarrow}=v_0\exp(W_\downarrow/k_BT)$ and $a_{\downarrow\uparrow}=v_0\exp(W_\uparrow/k_BT)$ are the transition probabilities at the temperature $T$.
\begin{center}
   \begin{figure}[ht]
    \centering$
        \begin{array}{c}
    \includegraphics[width=\columnwidth]{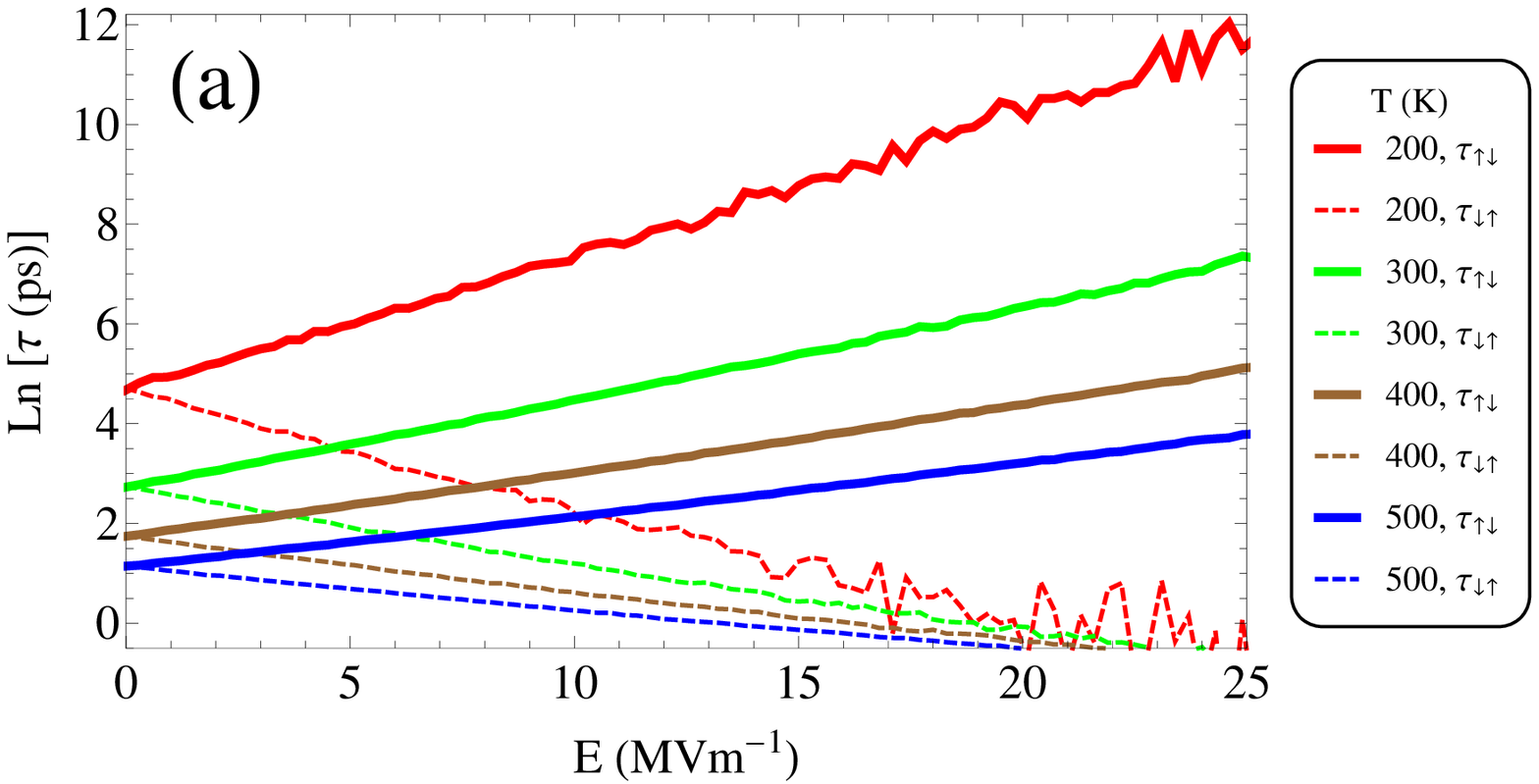}\\
    \includegraphics[width=\columnwidth]{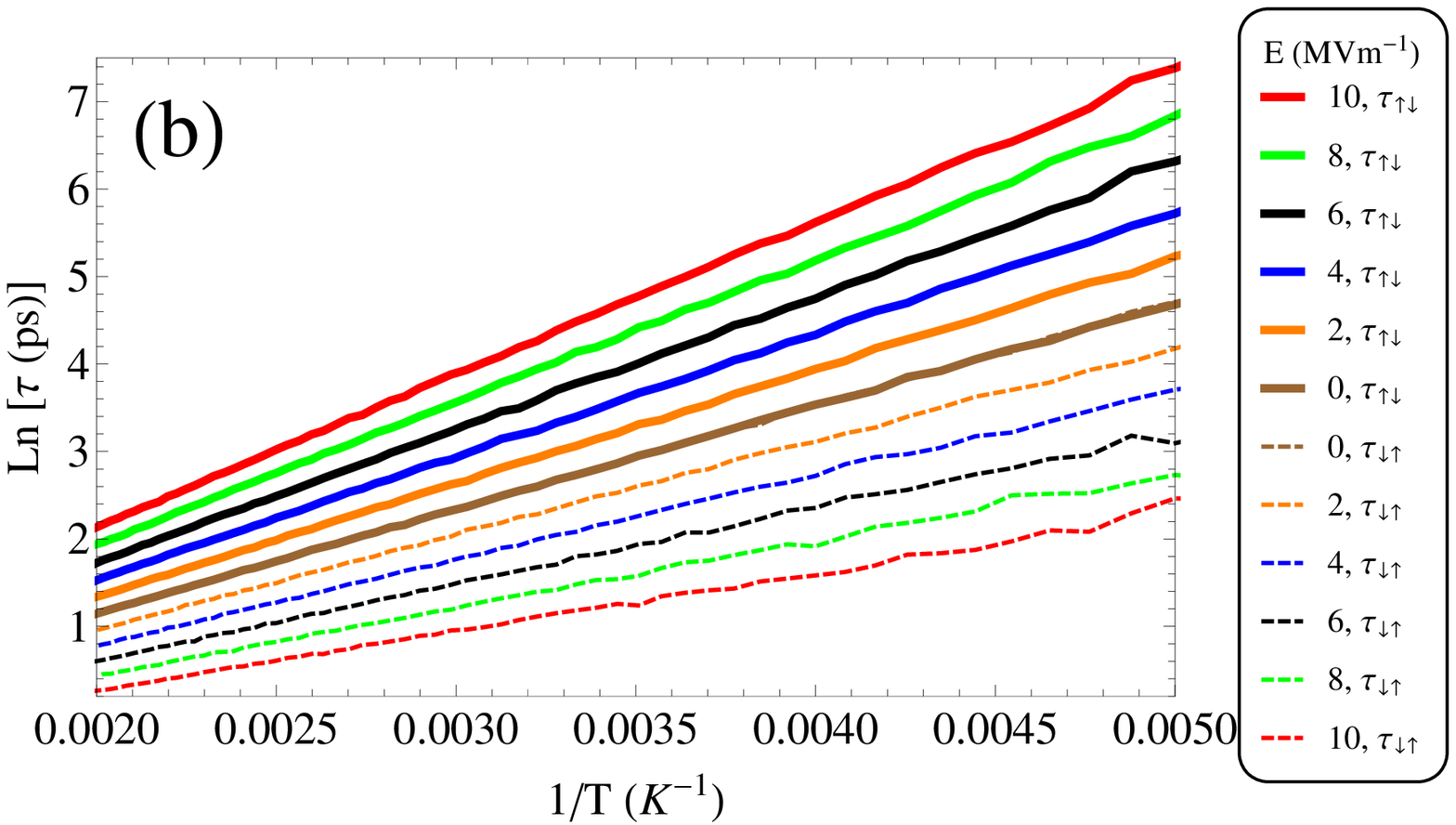}\\
        \end{array}$
        \caption{\label{t_sw} a) Thermally activated switching times as a function of the external electric field at different temperatures. b) Thermally activated switching times as a function of the temperature at different electric fields. $\tau_{\uparrow\downarrow}$ is the averaged time in which the polarization is in the up-state without switching (analogously, $\tau_{\downarrow\uparrow}$ for the state down, see Fig.~\ref{p_time}). The simulation time interval is fixed for all cases ($\Delta=1~\mu$s), therefore, a noisy behavior is observed as the switching time increases (either $\tau_{\uparrow\downarrow}$ or $\tau_{\downarrow\uparrow}$); $a^3=10^{-27}$ m$^3$ and $\gamma=25\times10^{-6}$ VmsC$^{-1}$.}
    \end{figure}
\end{center}

 \begin{center}
	\begin{figure}[b!]
		\centering$
		\begin{array}{c}
		\includegraphics[width=\columnwidth]{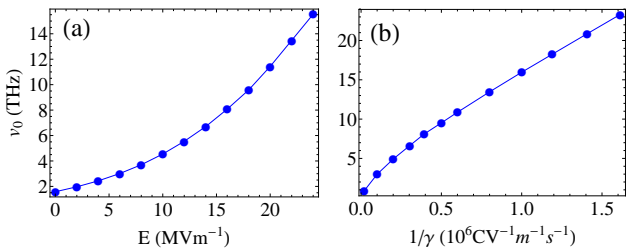}
		\end{array}$
		\caption{\label{v0-gama-E} a) Frequency of the optical phonons $v_0$ versus the electric field for $a^3=10^{-27}$ m$^3$ and $\gamma=25\times10^{-6}$ VmsC$^{-1}$. $v_0$ is evaluated by fitting the switching time formula (Eq.~(\ref{Switching_time_theory})) on the simulated switching times. For a better accuracy the Taylor expansion of $W_{\uparrow,\downarrow}$ up to O[E]$^5$ in Eq.~(\ref{Switching_time_theory}) is used. b) Frequency of the optical phonons $v_0$ versus the inverse internal resistivity at low temperatures for $a^3=10^{-27}$ m$^3$ and $E=0$.}
	\end{figure}
\end{center}

\subsection{Numerical framework}
To evaluate the switching time numerically using the Heun method we solve a Langevin type equation for the polarization dynamics averaging over individual switching times (see Fig.~\ref{p_time}). Here, the time-dependent Ginzburg-Landau (TDGL) equation is our Langevin type equation  \cite{Nambu}
\begin{equation}
\label{TDGL}
\begin{split}
\gamma\frac{\partial P}{\partial t}=-\frac{\partial F}{\partial P}+\eta(t),
\end{split}
\end{equation}
where $\gamma$ and $\eta$ are the internal resistivity (the inverse of kinetic coefficient), and a Gaussian white noise with the autocorrelation $\langle\eta(t)\eta(t')\rangle=q\gamma\delta(t-t')$, respectively (as follows from the fluctuation-dissipation theorem,  the strength of the noise must be related to $\gamma$). $q$ is unknown and we are going to show how it is related to the bath temperature.
To calculate $q$ the corresponding Fokker-Planck equation is required; we follow the Kramers-Moyal recipe given in Ref.~\cite{Risken} to obtain that. The drift and the diffusion coefficients are derived as $D^{(1)}=-\frac{\partial F}{\gamma\partial P}$ and $D^{(2)}=\frac{q}{2\gamma}$, respectively. By substitution of the drift and the diffusion coefficients in the Kramers-Moyal expansion the corresponding Fokker-Planck equation is derived
\begin{equation}
\label{FE_Fokker-Planck}
 \frac{\partial}{\partial t}w(P,t)=\frac{\partial}{\gamma\partial P}\left[w\frac{\partial F}{\partial P}+\frac{q}{2}\frac{\partial w}{\partial P}\right],
\end{equation}

where $w(P,t)dP$ denotes the differential probability of finding the polarization between the states $P$ and $P+dP$. Equivalently, we can follow a simpler intuitive method (similar to the case of a single ferromagnetic domain given in Ref. \cite{Brown}) to obtain the Fokker-Planck equation.
 We can start from the continuity equation $\frac{\partial}{\partial t}w(P,t)=-\frac{\partial}{\partial P} J$, where $J=wv-k'\frac{\partial w}{\partial P}$ is the probability current. $v=\frac{\partial P}{\partial t}=-\frac{\partial F}{\gamma\partial P}$ is the velocity of the representative point ($P$) along the polarization axis in the absence of thermal fluctuations and $-k'\frac{\partial w}{\partial P}$ is the diffusion term due to thermal fluctuations. Having the Fokker-Planck equation, we impose the requirement that in statistical equilibrium ($\partial w/\partial t=0$) $w$ must reduce to a Boltzmann distribution $w=w_0\exp(-a^3F/k_BT)$, then we find $q=\frac{2k_BT}{a^3}$ (or $k'=\frac{k_BT}{\gamma a^3}$)\cite{Marton,Nambu,SiWi04}. As expected $q$ represents the competition between the thermal energy and the potential barrier which itself scales as the volume $a^3$.

The results of the simulated polarization dynamic  is shown in Fig.~\ref{p_time}. Due to the stochastic nature of switching at finite temperatures we should perform the average  as to reach  convergence, while choosing the simulation time interval ($\Delta$) to be larger than the individual switching times ($\Delta\gg\tau$).

\subsection{Fokker-Planck vs. Pauli master equation}
In Fig.~\ref{t_sw} the switching times versus the electric field and the temperature for a sufficiently large simulation time interval ($\Delta=1~\mu$s $\gg\tau$) are shown. As expected, increasing the temperature decreases the switching times and the external electric field is (is not) in favor of $\tau_{\uparrow\downarrow}$ ($\tau_{\downarrow\uparrow}$). By fitting the data at $E=0$ to $\ln(\tau)=-\ln(2v_0)+\frac{W}{k_BT}$ we infer from  Fig.~\ref{t_sw}b  that  $W\approx-1.63\times10^{-20}$ J which is in a good agreement with the theoretical value of the minimum energy $W=-a^3\frac{\alpha^2}{4\beta}\approx-1.59\times10^{-20}$ J. Moreover, we can evaluate the frequency of the optical phonons as $v_0\approx1.68$ THz.
\begin{center}
   \begin{figure}[t!]
    \centering$
        \begin{array}{c}
    \includegraphics[width=\columnwidth]{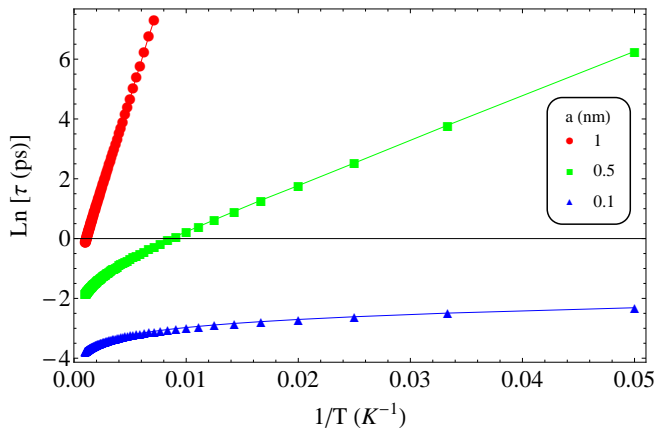}
        \end{array}$
        \caption{\label{gama} \textbf{Nonlinearity in reduced sizes :} Thermally activated switching times as a function of the temperature for different values of the cell sizes at $E=0$, and $\gamma=25\times10^{-6}$ VmsC$^{-1}$. Deviation from the theoretical linear formula $\ln(\tau)=-\ln(2v_0)+\frac{a^3\alpha^2/4\beta}{k_BT}$ is observed  for small volumes at high temperatures.}
    \end{figure}
\end{center}
Evaluating $v_0$ for different electric fields, Fig.~\ref{v0-gama-E}a shows that there is an increase in the frequency of the optical phonons by increasing the electric field. Beside the experimental evidence \cite{Worlock}, the dependency of $v_0$ on the electric field is understandable since it presents the total number of trials per second to overcome the potential barrier which itself depends on the electric field. Increasing  the electric field decreases the potential barrier and  increases so $v_0$. The optical phonon frequency is expected to be related to the internal resistivity ($\gamma$) as  there is no direct correspondence to the internal resistivity in Pauli master equation. Conceptually, $\gamma$ shows the resistivity of FEs to release charges by relinquishing the polarization (with an electric current density of $\frac{\partial P}{\partial t}$ (Eq.~(\ref{TDGL}))), which is related to the vibration of anions and cations in the FE atomistic structures. In other words, $\gamma$ determines the resistivity of FE to transfer from one state to another state. In this sense the inverse relation between the optical phonon frequency and the internal resistivity is comprehensible,  as shown in Fig.~\ref{v0-gama-E}b. Experimentally the internal resistivity is derived from Raman spectroscopy measurements, i.e. from measuring the
vibrational modes in ferroelectrics, and thus from the phonon frequencies \cite{Tenne}.

In summary, in addition to the relation between thermally activated switching time, the temperature, and the electric field in Fig.~\ref{t_sw}, by fitting the simulation data to the switching time formula (Eq.~(\ref{Switching_time_theory})) we arrived at a relation between $v_0$, and electric field, and the internal resistivity in Fig.~\ref{v0-gama-E}. This endorses that  introducing the temperature as a thermal noise into FEs is useful. However, it seems that in some circumstances our simulated switching times deviate from the theoretical formula. In the absence of an electric field, according to the Eq.~(\ref{Switching_time_theory}) the relation between $\ln(\tau)$ and $\frac{1}{T}$ \emph{must} be linear. But as can be seen in Fig.~\ref{gama}, at reduced sizes, as the temperature rises a nonlinear regime appears. We attribute this nonlinearity to the approximation of small thermal energies ($k_{\mathrm{B}}T<<W$) assumed during the derivation of Eq. (\ref{Switching_time_theory}).
We will elaborate on the size effects in FEs to show the importance of revising  numerical treatments at finite temperatures.

\begin{center}
	\begin{figure}[t!]
		\centering$
		\begin{array}{c}
        \includegraphics[width=\columnwidth]{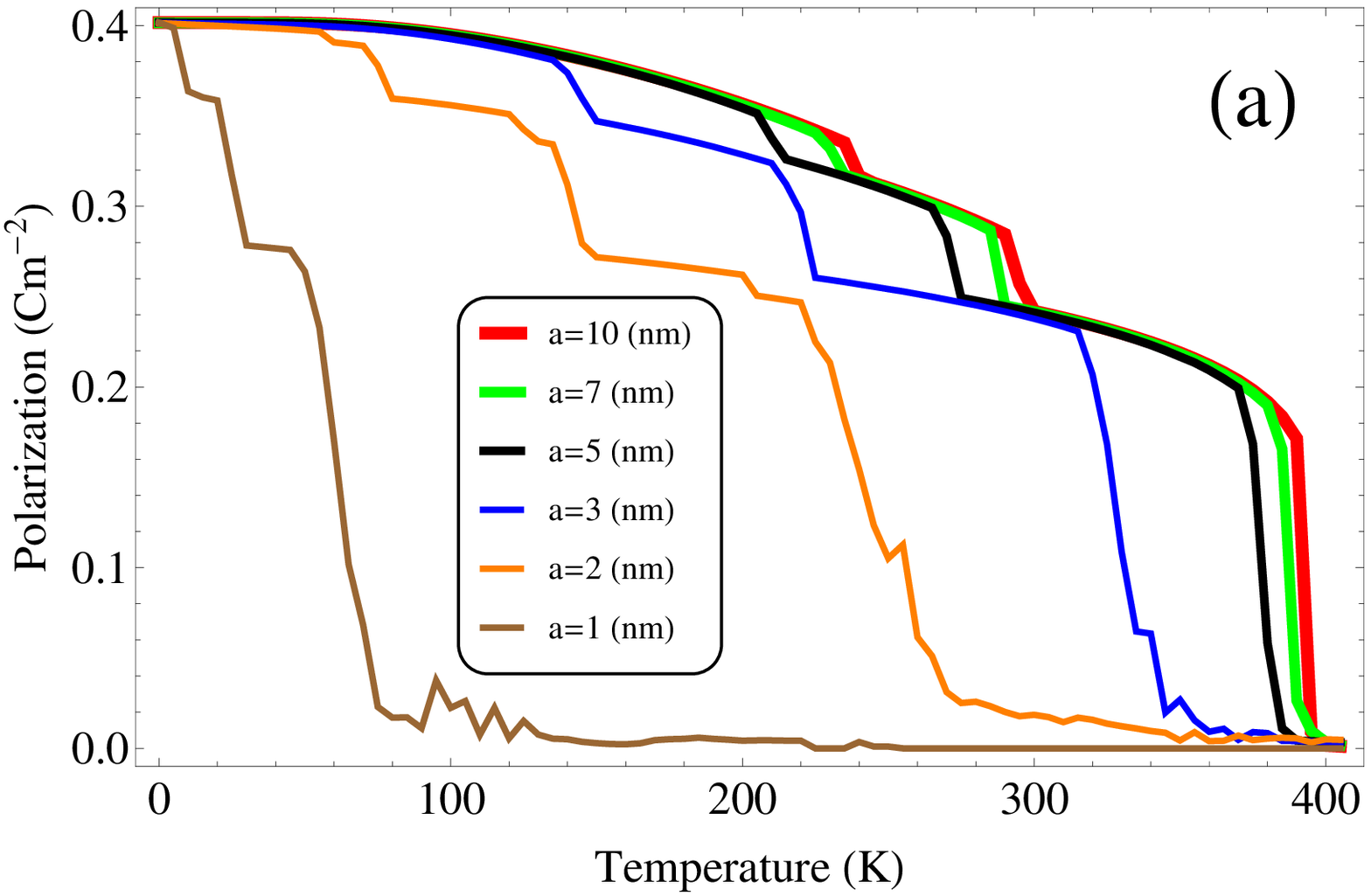}\\
        \includegraphics[width=\columnwidth]{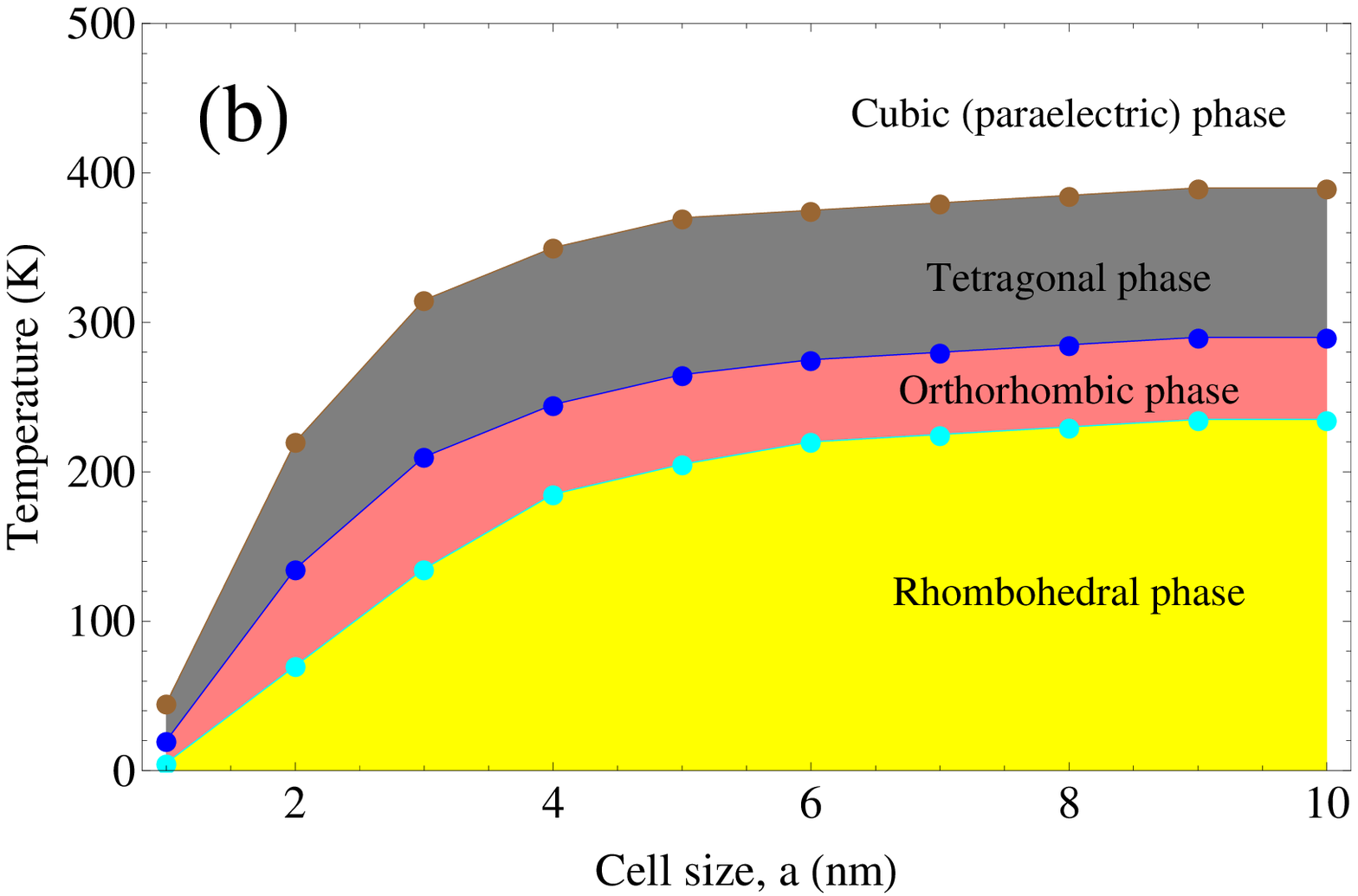}
		\end{array}$
		\caption{\label{phase_diagram} \textbf{Phase instability at reduced sizes :} a) Averaged total polarization (over 50 realizations of the noises) for a single domain BTO with different cell sizes. For large cell sizes, a phase diagram in line with the experimental one is obtained.
The jumps correspond to the structural phase transitions. b) The corresponding structural phase transition temperatures are shown. For more details see the Appendix.}
	\end{figure}
\end{center}

\section{Phase instability}
The investigation of finite-size effects in FEs has experienced a great impetus due to miniaturization of electronic devices. We address the key feature of  the suppression of polarization at reduced sizes \cite{Sedykh,LiEastman1996,Hoshina,Ishikawa,Lichtensteiger,Junquera,Lee,Scott,Zhao} which  hampers  functionalities, e.g.
 high-density memory applications. We focus on the prototypical ferroelectric barium titanate (BTO) which has already been used in electronic devices such as multilayered ceramic capacitors.  We use the eighth-order Landau-Devonshire potential to evaluate its phase diagram
\begin{equation}
\label{potential}
\begin{split}
 F_{\mathrm{GLD}}&=\alpha_1(T)(P_{x}^2+P_{y}^2+P_{z}^2)+\alpha_{11}(P_{x}^4+P_{y}^4+P_{z}^4)\\
 &+\alpha_{12}(P_{x}^2P_{y}^2+P_{y}^2P_{z}^2+P_{x}^2P_{z}^2)+\alpha_{111}(P_{x}^6+P_{y}^6+P_{z}^6)\\
 &+\alpha_{112}[P_{x}^2(P_{y}^4+P_{z}^4)+P_{y}^2(P_{x}^4+P_{z}^4)+P_{z}^2(P_{x}^4+P_{y}^4)]\\
 &+\alpha_{123}P_{x}^2P_{y}^2P_{z}^2+\alpha_{1111}(P_{x}^8+P_{y}^8+P_{z}^8)\\
 &+\alpha_{1112}[P_{x}^6(P_{y}^2+P_{z}^2)+P_{y}^6(P_{x}^2+P_{z}^2)+P_{z}^6(P_{x}^2+P_{y}^2)]\\
 &+\alpha_{1122}(P_{x}^4P_{y}^4+P_{y}^4P_{z}^4+P_{x}^4P_{z}^4)\\
 &+\alpha_{1123}(P_{x}^4P_{y}^2P_{z}^2+P_{y}^4P_{z}^2P_{x}^2+P_{z}^4P_{x}^2P_{y}^2),
\end{split}
\end{equation}
which has already been used and tested in the literature \cite{Wang1,Li,Wang}.
All the parameters are taken from Ref.~\cite{Wang1}. The important point, to which we would like to draw  attention to, is that in  previous works the temperature was introduced  solely via the potential coefficients.
At first glance, the model seems to lack any finite-size effects.
 Here along with the temperature-dependent potential coefficients, we introduce the temperature as a thermal noise, as well (similar to Eq.~(\ref{TDGL}) but for all three components of the polarization). This is important insofar as when the size of the system decreases,   the thermal fluctuations activate the order parameter
 to overcome the potential barrier, and finite-size effects play a role. As evident from Fig.~\ref{phase_diagram}, including the thermal noise gives a clear finite-size effect. On the one hand, as the size of the FE domain decreases the polarization is suppressed and also a phase instability appears. On the other hand, as the size of FE domain increases the phase diagram approaches the well-established macroscopic one \cite{Wang1}.

\begin{center}
	\begin{figure}[t!]
		\centering$
		\begin{array}{cc}
		\includegraphics[width=\columnwidth]{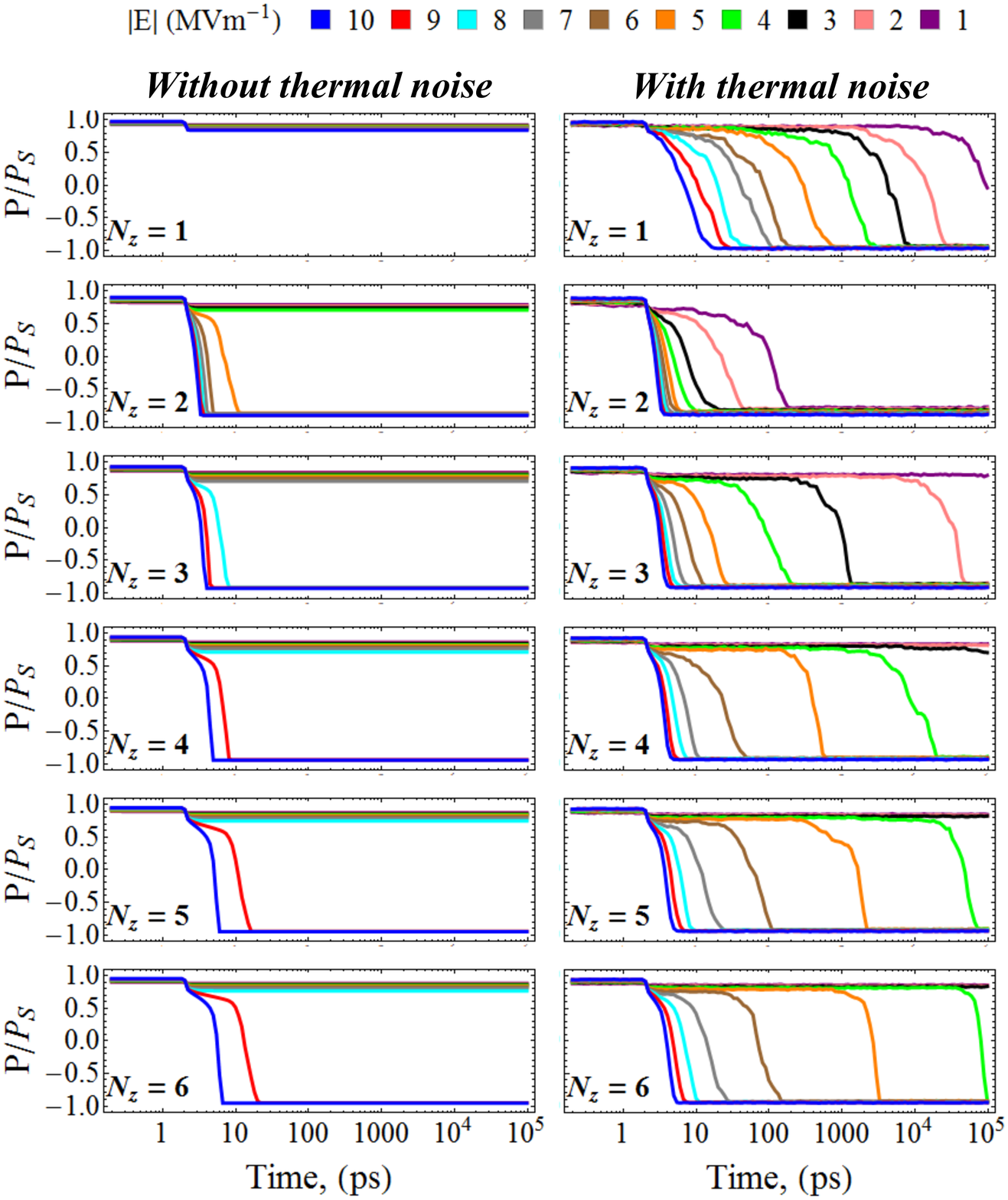}
		\end{array}$
		\caption{\label{Electric-field_activated_sw} \textbf{Electric-field activated switching time.} The trajectory of the polarization for $10\times10\times N_{\mathrm{z}}$ coupled BTO domains with $a=3$ nm at room temperature is shown. The switching of the polarization occurs due to the switching of the electric field in contrast to the thermally activated switching time (Fig.~\ref{p_time}) where switching was due to thermal fluctuations   only  . However, here the temperature could facilitate the switching and as can be seen it is relevant how the temperature is introduced into the system. Up to $2$ ps the electric field is along the $+z$ and later it abruptly switches to $-z$.}
	\end{figure}
\end{center}

\begin{center}
	\begin{figure}[t!]
		\centering$
		\begin{array}{c}
		\includegraphics[width=\columnwidth]{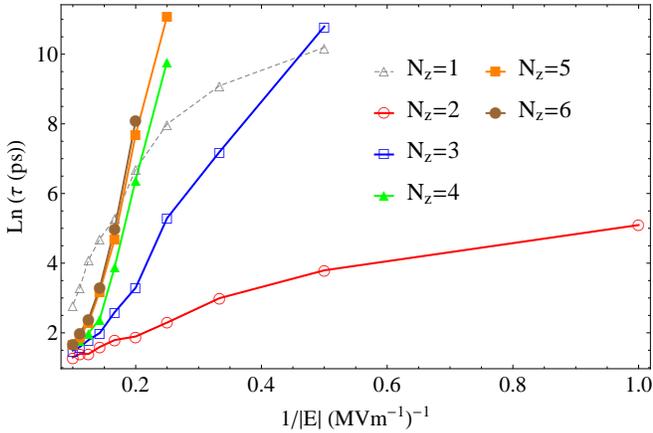}
		\end{array}$
		\caption{\label{Electric-field_activated_sw2} The electric-field activated switching times versus the inverse electric field for different thicknesses of a $10\times10\times N_{\mathrm{z}}$ coupled BTO domains with $a=3$ nm at room temperature is shown. The temperature is not only introduced via the potential coefficients but also via the thermal noises (Fig.~\ref{Electric-field_activated_sw}).}
	\end{figure}
\end{center}

\begin{center}
	\begin{figure*}[t!]
		\centering$
		\begin{array}{c}
		\includegraphics[width=17cm]{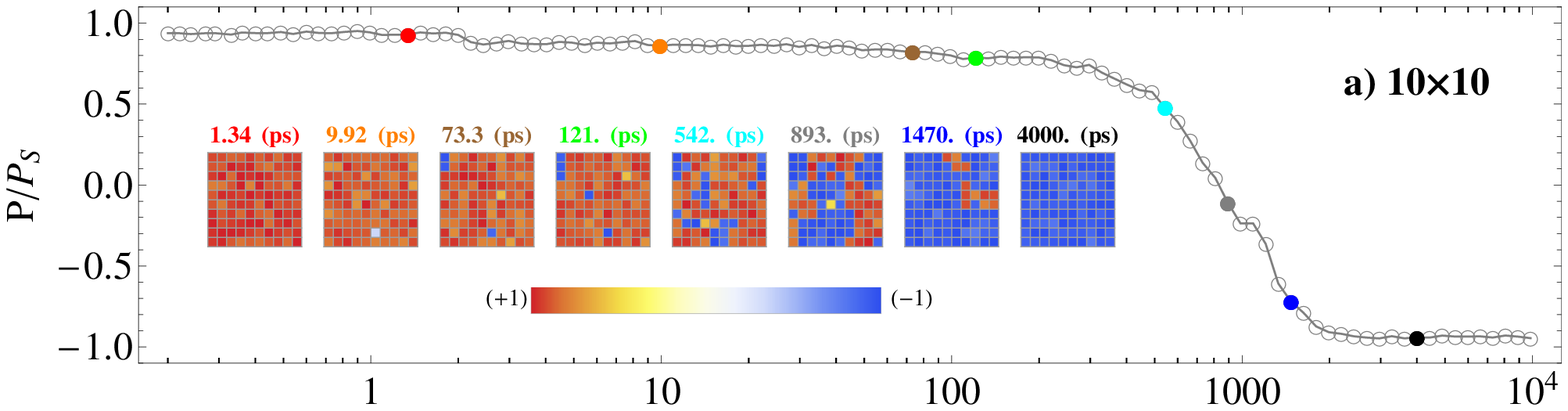}\\
		\includegraphics[width=17cm]{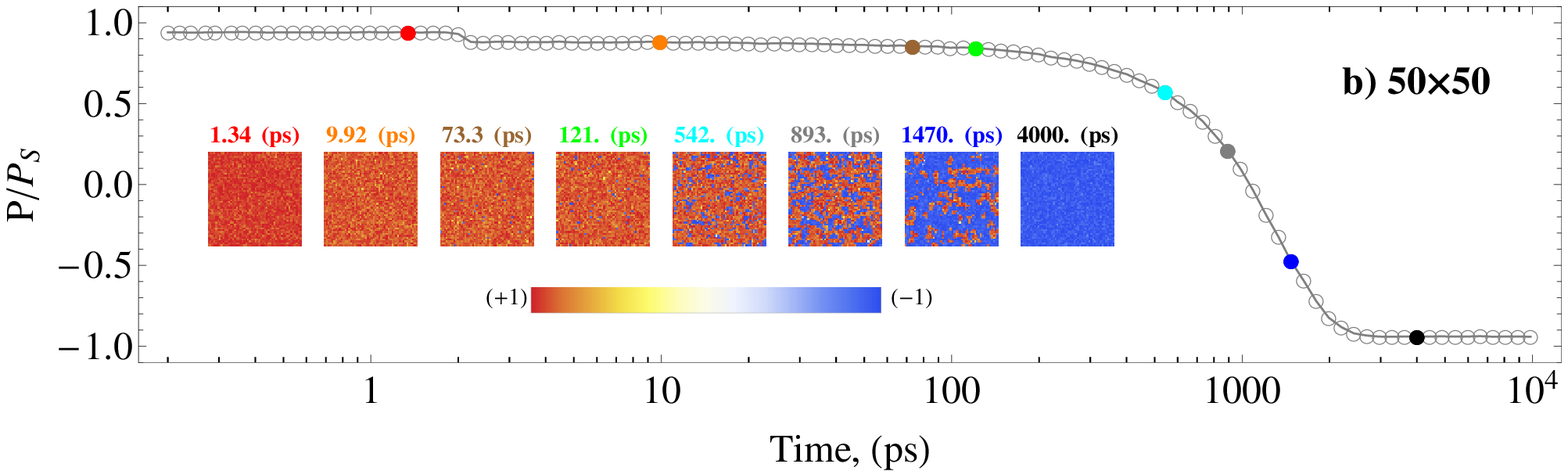}
		\end{array}$
		\caption{\label{Electric-field_activated_sw3} \textbf{Electric-field activated switching time.} a) The figure shows the trajectory of the total polarization for $10\times10\times1$ coupled BTO domains with $a=3$ nm at room temperature. Up to $2$ ps the electric field (4 MVm$^{-1}$) is along the $+z$ and later it abruptly switches to $-z$. The insets show the state of the polarization of domains at each of the depicted times. The nucleation and the sideways-domain wall motion mechanisms are evident. b) The same, but for $50\times50\times1$ coupled BTO domains. The expansion of the system in $x-y$ plane does not lead to considerable size effects.}
	\end{figure*}
\end{center}

\section{Electric-field activated switching time in multi-domain systems}
\begin{center}
	\begin{figure}[t!]
		\centering$
		\begin{array}{c}
		\includegraphics[width=\columnwidth]{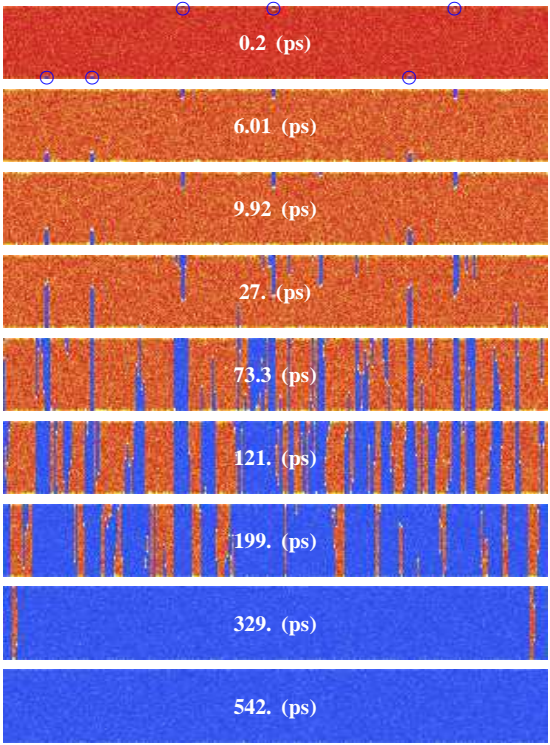}
		\end{array}$
		\caption{\label{pinning} The state of the polarization for $300\times1\times50$ coupled BTO domains ($N_{\mathrm{x}}=300$, $N_{\mathrm{y}}=1$ and $N_{\mathrm{z}}=50$) with $a=3$ nm at room temperature at different times are shown. Up to $2$ ps the electric field (6 MVm$^{-1}$) is along the $+z$ and later it abruptly switches to $-z$ (the same as the other simulations, it is assumed that electrodes are highly conductive : $E_{dep}\approx0$). Circles show the pinning sites. The nucleation starts from the pining sites and later on from some random sites at the top and bottom of the sample. After that the reversal occurs by forward growth and finally sideways growth of the domains \cite{Dawber}.}
	\end{figure}
\end{center}

Finally, we simulate the electric-field activated switching time for coupled BTO domains at room temperature to provide another evidence for the necessity of including   thermal noise along with the temperature-dependent potential coefficients into the models. To address nonhomogeneous multi-domain state the Ginzburg gradient is added to the Landau-Devonshire potential \cite{Hlinka,Hlinka1,Marton1}.   At room temperatures, in which case  BTO is in the tetragonal phase, the polarization has two stable orientations. The total free energy density reads $F_{\mathrm{tot}}=F_{\mathrm{GLD}} + F_{\mathrm{G}} - \sum_\textbf{n} \vec{P}_\textbf{n} \cdot \vec{E}$, where the expression for the gradient energy density $F_{\mathrm{G}}$ reads
\begin{equation}
\label{tetragonal}
\begin{split}
F_{\mathrm{G}}=\sum_{\textbf{n}}&\frac{G_{44}}{2a^2}(P_{(n_x+1,n_y,n_z)}-P_{(n_x,n_y,n_z)})^2\\
+&\frac{G_{44}}{2a^2}(P_{(n_x,n_y+1,n_z)}-P_{(n_x,n_y,n_z)})^2\\
+&\frac{G_{11}}{2a^2}(P_{(n_x,n_y,n_z+1)}-P_{(n_x,n_y,n_z)})^2;
\end{split}
\end{equation}
where $G_{44}=2\times10^{-11}$ [C$^{-2}$m$^3$J] and $G_{11}=51\times10^{-11}$ [C$^{-2}$m$^3$J] determine the strength of the coupling between the BTO domains \cite{Hlinka}. We consider a 3D system including $10\times10\times N_{\mathrm{z}}$ domains which $N_{\mathrm{z}}$   determines the thickness of the system. The domain size is chosen above $a\sim2.9$ nm to assure a nonzero spontaneous polarization at each domain at room temperature (see Fig.~\ref{phase_diagram}). The polarization dynamic versus the time for $a=3$ nm and for different thicknesses are shown in Fig.~\ref{Electric-field_activated_sw}. This figure also shows the influence of the thermal noise. Our simulations  confirm clearly  that introducing the temperature as a thermal noise, in addition to the temperature-dependent potential coefficients, is required to model the experimental observations \cite{LiAl07} (at least qualitatively). Moreover, according to the Merz's law \cite{Merz54,Merz_1956,Viehland2000,Viehland2001,Lou}, it is expected that at low electric fields the electric-field activated switching time obeys the following relation
\begin{equation}
\label{tetragonal}
\begin{split}
\tau\sim\exp\left(\lambda/|E|\right),
\end{split}
\end{equation}
which is a special case of the KAI model and is confirmed by our simulations (Fig.~\ref{Electric-field_activated_sw2}). The exponential behavior can be  understood statistically in terms of the nucleation of random sites \cite{Fatuzzo_1959,Fatuzzo_1962}, though other mechanisms also involve in the switching (Fig.~\ref{Electric-field_activated_sw3}). We attribute the deviation from the exponential behavior at large electric fields to the domination of domain-wall motion mechanism over the nucleation mechanism, as mentioned by Merz \cite{Merz_1956}. The anomalous change from $N_{\mathrm{z}}=1$ to $N_{\mathrm{z}}=2$ is the result of the strong coupling term along the z axis being added to the free energy.
For $N_{\mathrm{z}}>1$ the forward-domain wall motion mechanism is incorporated in  the nucleation and sideways-domain wall motion mechanisms (Fig.~\ref{pinning}). $\lambda$ is known as the activation field and its value critically depends on the experimental factors. Experimentally, the polarization switching requires two electrodes for applying the electric field. Their sole presence affects the switching. In the electrodes, the bound surface charges, that are arising from the discontinuous normal polarization component at the interface, are compensated by the screening with free charges in the electrodes immediately at the interface. Real electrodes are not ideal conductors and leave a small depolarization field because of the finite screening length that characterizes the space-charge extent in the electrodes. Without electrodes, depolarization field may well be large. The depolarization field can be estimated as $E_{dep}=-\frac{P_S}{\epsilon_F}\left(\frac{2\epsilon_F/l}{2\epsilon_F/l+\epsilon_e/l_s}\right)$ where $\epsilon_F$, $\epsilon_e$, $l$ and $l_s$ are the dielectric constant of the FE, the dielectric constant of the electrodes, the thickness of the FE and the screening length, respectively \cite{Mehta_1973}. To uncover the sole role of thermal fluctuations, within all the simulations we assumed the FE sample is attached to highly conducting electrodes ($\epsilon_e/l_s\rightarrow\infty$), meaning $E_{dep}\approx0$ \cite{Merz_1956,kanzig_1955}. Moreover, we have assumed a defect free FE material. Numerically, however, we can keep the polarization of some sites fixed (as the pinning sites) apart from the dynamics of the remainder of FE material to explore the role of  defects  (Fig.~\ref{pinning}) \cite{PiLe04,Jo_2009}. Our focus on this study is on the importance of thermal fluctuations in assisting the polarization switching, a deeper investigation of this important  issue of switching in general  \cite{Ducharme_2000,Landauer_1957,Jo_2006,Highland_2010,Tian_2015,Baudry_2015,Baudry_2015,Liu_2016,Shin_2007,Guo_2015,Gaynutdinov_2013,Lou_2009,Jiang_2009,Stolichnov_2004,Ivry_2012}, within our formalism deserves further investigations and analysis.

\begin{center}
	\begin{figure}[t!]
		\centering$
		\begin{array}{c}
		\includegraphics[width=\columnwidth]{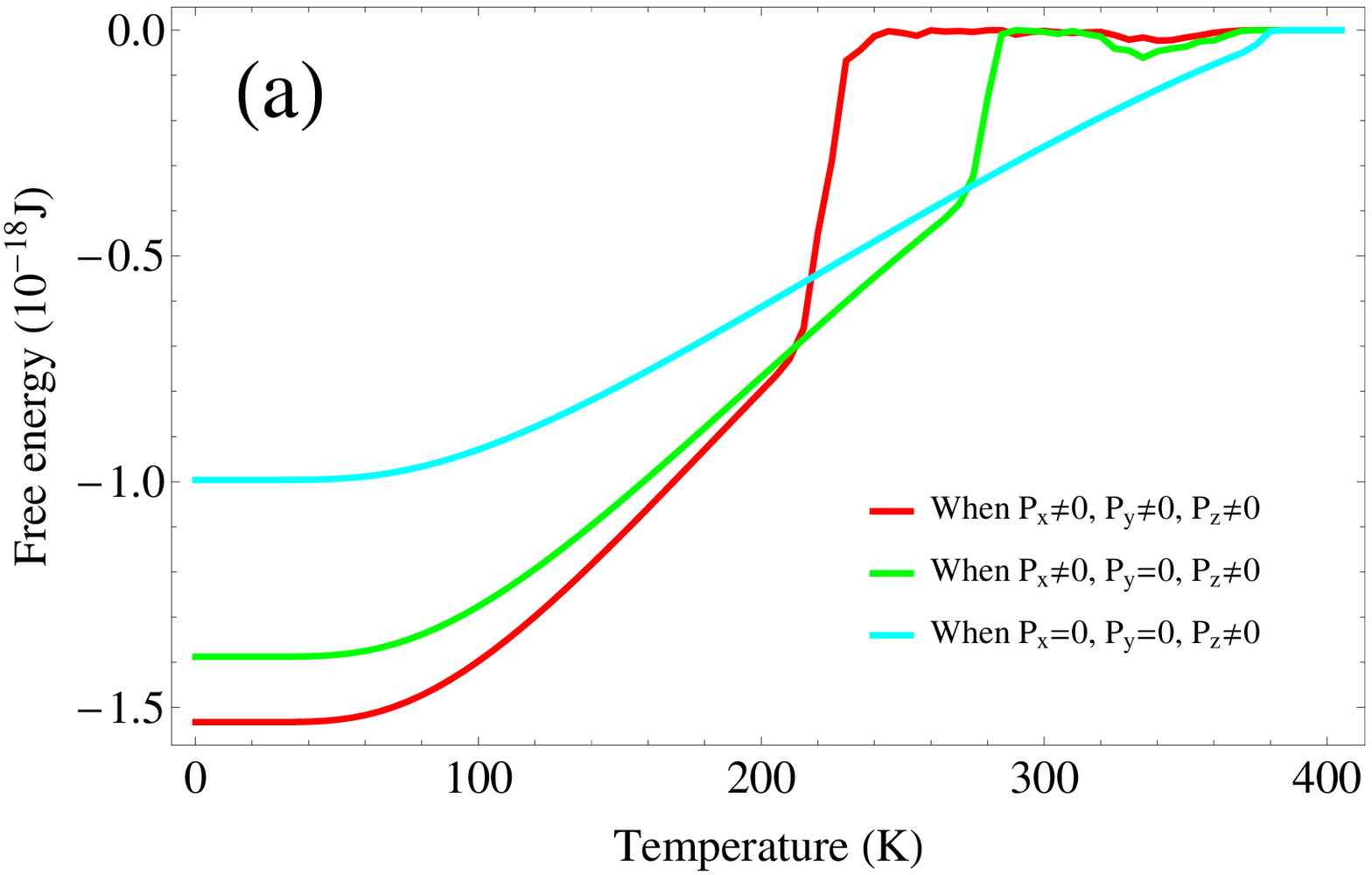}\\
		\includegraphics[width=\columnwidth]{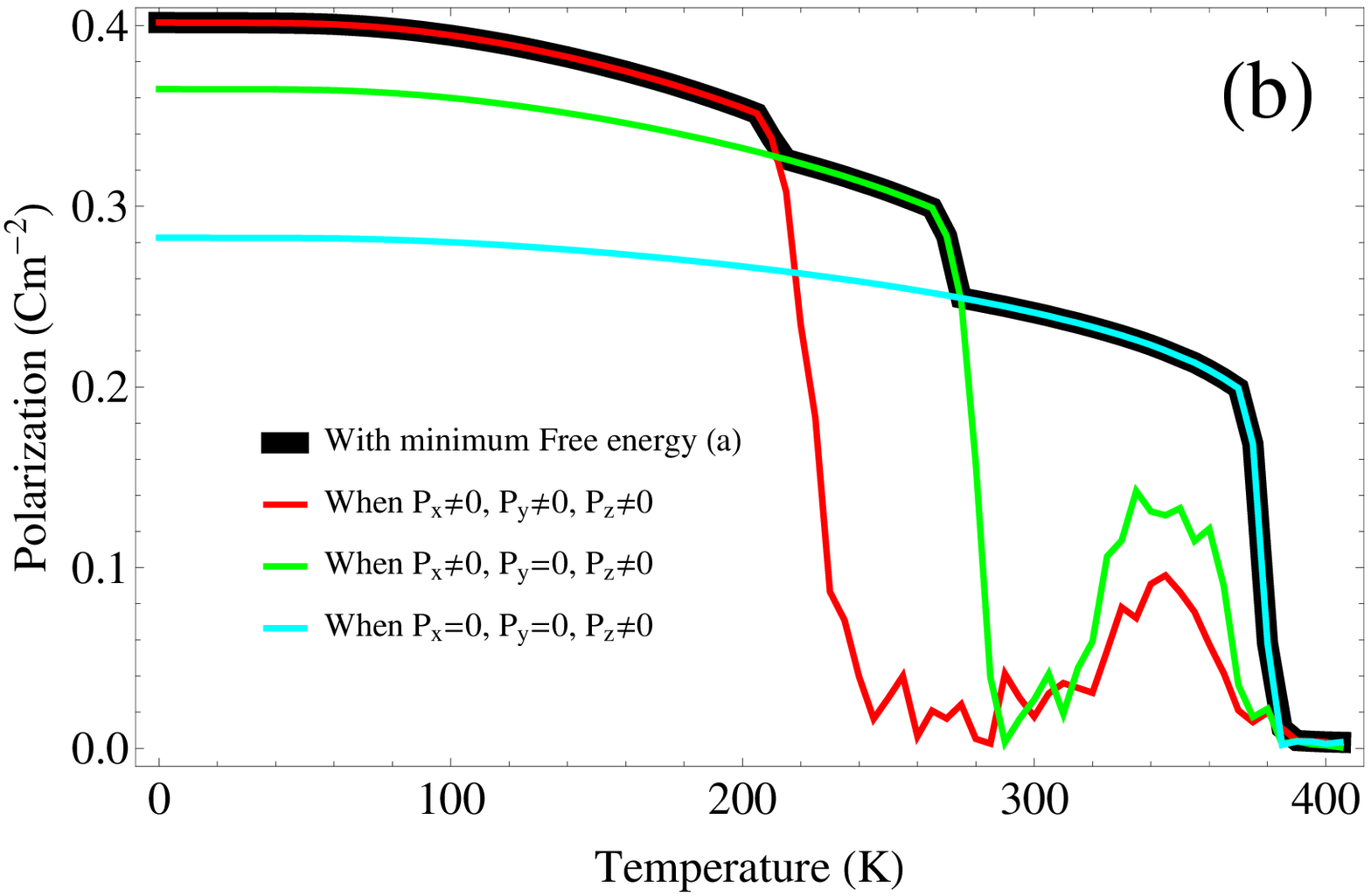}\\
		\includegraphics[width=\columnwidth]{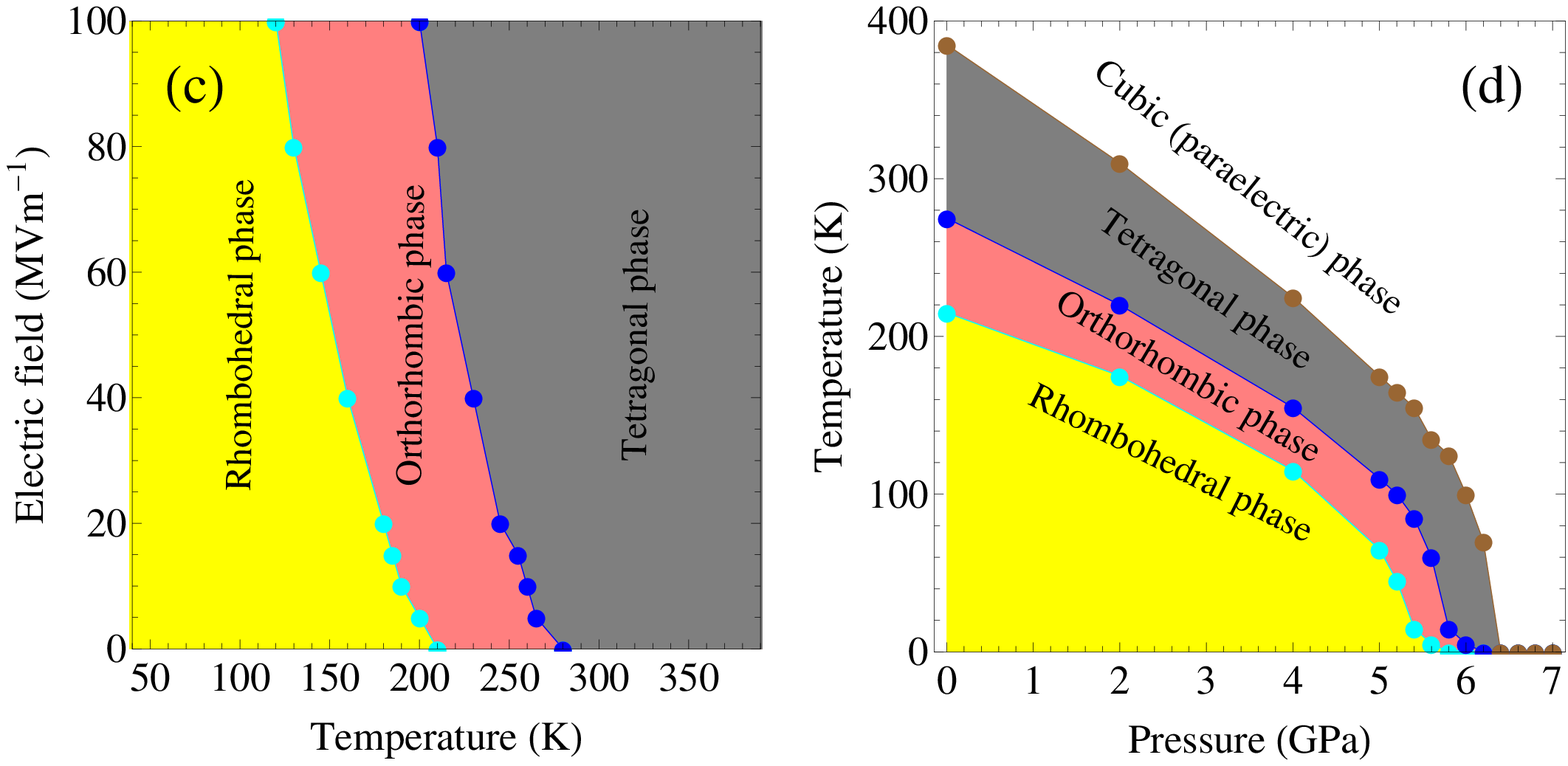}
		\end{array}$
		\caption{\label{FE_self_consistense_TC} \textbf{Phase diagram calculation recipe.}
 a) shows the free energies for three different cases  corresponding to three different structural phases at zero electric field and zero hydrostatic pressure for a single BTO domain with $a=5$ nm. b) shows the  polarization $P=\sqrt{P_x^2+P_y^2+P_z^2}$ corresponding to three different structural phases.
At each temperature the polarization that corresponds to the minimum free energy  represents the correct polarization which is marked by a thick black curve. c)-d) Following the same recipe as a) and b) the structural phase transitions for different electric fields and hydrostatic pressures are calculated using the extended potential given in Ref.~\cite{Wang1}.}
	\end{figure}
\end{center}

\section{Summary and Conclusions}
Using a simple model we compared two different theoretical frameworks for thermally activated switching times in FEs. The first framework was based on the Pauli master equations which provided an expression for the switching time (Eq. (\ref{Switching_time_theory})), while the second framework was based on the Fokker-Planck formalism and  serves as a basis for our simulations (Eq. (\ref{TDGL})). We evaluated the optical phonon frequency as an outcome of the comparison and found that the two frameworks are consistent within the range of the used parameters (Fig.~\ref{v0-gama-E}). At reduced sizes and elevated temperatures, however, a nonlinear regime emerged in our simulations (Fig.~\ref{gama}), which was related to the violation of the approximation of Eq. (\ref{Switching_time_theory}) of low T-fluctuations with respect to the FE energy barrier. We calculated the phase diagram for different sample sizes for BTO that revealed the phase instability at reduced sizes as frequently reported in the literature \cite{Sedykh,LiEastman1996,Hoshina,Ishikawa,Lichtensteiger,Junquera,Lee,Scott,Zhao}. We found that at reduced sizes thermal fluctuations are sizable and are on the scale of the internal FE potential and can suppress the polarization (Fig.~\ref{phase_diagram}); which is reminiscent to  the superparaelectric behavior in the relaxor ferroelectrics \cite{Cross,Skulski,Tiruvalam,Bokov,Li2,Westphal} and to the theoretically predicated superparaelectric phase in  ferroelectric nanoparticles \cite{Glinchuk}.
Finally, to further illustrate the importance of the thermal noise for FE dynamics, we simulated the electric-field activated switching times for multi-domain BTO samples. If  the thermal noise is present we obtained  results in accordance with the KAI model (Figs.~\ref{Electric-field_activated_sw} and \ref{Electric-field_activated_sw2}). \\
Our results show that  generally for simulating the FE dynamics,  the  temperature must be incorporated   \emph{both} via a noise term   and potential coefficients.
 Notably,  a similar mathematical model based on the solution of the stochastic Landau-Lifshitz-Bloch equation \cite{Gara97,Gara04,Atxi07,Kaza07,Evan12}, which accounts for both the reduction of the order parameter (magnetization for this matter) at temperatures close to the transition temperature and the fluctuation of the order parameter due to thermal noise, provides a viable method for  describing  the magnetization switching in the emerging field of heat-assisted magnetization recording. Significant differences between the polarization and the magnetization dynamics at temperatures near the phase transition concern the fact that in FEs thermal fluctuations affect mainly the longitudinal component of the polarization, whereas for ferromagnets the fluctuations are in both the longitudinal and the transversal components of the magnetic moment, but mainly transversal at moderate temperatures.

\section{Acknowledgment}
The authors would like to thank Kathrin D\"orr for valuable discussions and suggestion on the experimental aspects of our study.

\section{Appendix}
Here we show in brief how the BTO phase diagram including structural transitions is calculated. To achieve this, at first we need to calculate the free energy of the system and  the corresponding polarization at the stationary state for three different cases. Fig.~\ref{FE_self_consistense_TC}a  corresponds to the three different structural phases. At each temperature among the threes phases, we choose a polarization which its associated with the free energy  minimum. With this recipe, the transition temperatures are obtained self-consistently. The Fig.~\ref{FE_self_consistense_TC}b is just for a zero electric field and a zero hydrostatic pressure. We can perform the same recipe for different electric fields (Fig.~\ref{FE_self_consistense_TC}c), hydrostatic pressures (Fig.~\ref{FE_self_consistense_TC}d) and cell sizes (Fig.~\ref{phase_diagram}b) to obtain the corresponding structural phase transitions.

\end{document}